\documentclass[reprint,floatfix,nofootinbib]{revtex4-1}

\usepackage{amsfonts}
\usepackage{amsmath}
\usepackage{amssymb}
\usepackage{graphicx}
\usepackage{times}
\usepackage{hyperref}
\usepackage{epsfig}
\usepackage{dcolumn}
\usepackage{epstopdf}
\usepackage{circuitikz}
\usepackage{tikz}

\newcommand{\bear}{\begin{eqnarray}}
\newcommand{\eear}{\end{eqnarray}}
\newcommand{\be}{\begin{equation}}
\newcommand{\ee}{\end{equation}}
\newcommand{\Eqref}[1]{Eq.~(\ref{#1})}

\newcommand{\nn}{\nonumber}

\begin{document}

\title{Simple do-it-yourself experimental set-up for electron charge $q_e$ measurement}

\author{T.~M.~Mishonov}
\email[E-mail: ]{mishonov@gmail.com}
\affiliation{Laboratory for Measurements of Fundamental Constants,
Faculty of Physics,\\
St.~Clement of Ohrid University at Sofia,\\
5 James Bourchier Blvd., BG-1164 Sofia, Bulgaria}

\author{E.~G.~Petkov}
\affiliation{Laboratory for Measurements of Fundamental Constants,
Faculty of Physics,\\
St.~Clement of Ohrid University at Sofia,\\
5 James Bourchier Blvd., BG-1164 Sofia, Bulgaria}

\author{N.~Zh.~Mihailova}
\affiliation{Laboratory for Measurements of Fundamental Constants,
Faculty of Physics,\\
St.~Clement of Ohrid University at Sofia,\\
5 James Bourchier Blvd., BG-1164 Sofia, Bulgaria}

\author{A.~A.~Stefanov}
\affiliation{Faculty of Mathematics,\\
St.~Clement of Ohrid University at Sofia,\\
5 James Bourchier Blvd., BG-1164 Sofia, Bulgaria}
\affiliation{Institute of Mathematics and Informatics, Bulgarian Academy of Sciences, \\
Acad. Georgi Bonchev Str., Block 8, 1113 Sofia, Bulgaria}

\author{I.~M.~Dimitrova}
\affiliation{Faculty of Chemical Technologies, University of Chemical Technology and Metallurgy,\\
8 Kliment Ohridski blvd, BG-1756 Sofia, Bulgaria}

\author{V.~N.~Gourev}
\affiliation{Department of Atomic Physics, Faculty of Physics,\\St.~Clement of Ohrid University at Sofia,\\5 James Bourchier Blvd., BG-1164 Sofia, Bulgaria}

\author{N.~S.~Serafimov}
\affiliation{Department of Telecommunications,\\Technical University Sofia,\\8 St. Clement Blvd., BG-1000 Sofia, Bulgaria}

\author{V.~I.~Danchev}
\email[E-mail: ]{spectrefdx@gmail.com}
\affiliation{Department of Theoretical Physics, Faculty of Physics,\\
St.~Clement of Ohrid University at Sofia,\\
5 James Bourchier Blvd., BG-1164 Sofia, Bulgaria}

\author{A.~M.~Varonov}
\email[E-mail: ]{avaronov@phys.uni-sofia.bg}
\affiliation{Department of Theoretical Physics, Faculty of Physics,\\
St.~Clement of Ohrid University at Sofia,\\
5 James Bourchier Blvd., BG-1164 Sofia, Bulgaria}

\date{25 October 2018}

\begin{abstract}
A simple experiment for the electron charge $q_e$ measurement is described.
The experimental set-up contains standard electronic equipment only and can be built in every high-school lab all around the world with pocket money budget for several days.
It is concluded that it is time such a practice to be included in the regular high-school education.
The achieved 13\% accuracy is comparable to the best student university labs.
The measurement is based on Schottky noise generated by a photodiode.
Using the criterion dollar per accuracy for the electron charge $q_e$ measurement,
this definitely is the world's best educational experiment.
An industrial replica can be easily sold across the globe.
\end{abstract}

\maketitle

\section{Introduction}
\label{sec:intro}

The electron charge $q_e$ is included in the program of high-school education on physics.
Unfortunately, $q_e$ is not measured in high schools and even in many of the best universities
it is determined with poor accuracy using kilodollar (k\$) equipment.\footnote{\href{https://doi.org/10.1088/1361-6404/aad3d7}{T M Mishonov et al 2018 Eur. J. Phys. \textbf{39} 065202} \\  DOI: 10.1088/1361-6404/aad3d7}

Let us recall cursory the history of the experimental method.
Young Walter Schottky a student of Max Plank attended a lecture by Albert Einstein on fluctuations;
cf. his obituary Ref.~\onlinecite{Welker:76}.
Impressed that fluctoscopy can be used to determine fundamental constants, 
Schottky suggested determining the electron charge $q_e$ by investigating current fluctuations~\cite{Schottky:18,Schottky:26} exactly a century ago.
The Schottky relation $(I^2)_f = 2 q_e \left < I \right >$ between the
the averaged current $\left < I \right >$ and the spectral density of the current noise 
$(I^2)_f $ will be discussed later.
The experiment suggested by Schottky was performed as a research in the field of the fundamental physics~\cite{Hull:25}
and led to a precise determination of the electron charge~\cite{Stigmark:52}.

For physics education, reaching a metrological accuracy is not necessary 
and the accent is to describe experimental set-ups giving illustration of the physics principles.
Already for half a century the determination of $q_e$ by shot noise has been an eternal theme 
of articles addressed to the physics teachers and students laboratories, 
including a dozen articles~
\cite{Earl:66,Livesey:73,Kittel:78,Vetterling:79,Ericson:88,Kraftmakher:95,Spiegel:95,Cristofolini:03,
Kraftmakher:05,Rodriguez:10,Wash:12,MIT:13,Dessau:16},
most of which are in the American Journal of Physics (AJP) and are expensive for the high school physics (several k\$ range).
However, all those works give a contemporary version of the original set-up.
Even a recent work requires~\cite{Rice:16}:
high-level electronics controller, low-level
electronics controller, temperature module w/ probe, break-out
box, clear dewar in adjustable height support, coax cables,
45~W +/-15~V power supply, hook up wire, resistors,
transistors, diodes, photodiode in holder, light bulbs and
LEDS, and the commercial price of all this exceeds 7k\$. 
See also the Teach Spin setup~\cite{Noise_Signal}.
The purpose of this research is to describe a sufficiently different set-up, for which
all those expensive accessories are not necessary, while the accuracy is the same.

The experimental set-up contains standard electronic parts, which can be delivered
in every town within a week:
two low-noise dual operational amplifiers (OpAmps), one analog multiplier,
a prototype board, resistors, capacitors (four of them of film layer type), standard type batteries, an incandescent bulb (or white LED with luminophore), a potentiometer, a photodiode  and two standard multimeters. 
As out of school physics education and for an university freshman, 
the construction of the described set-up is a great starting point in electronics.
Every high school student rotating the axis of the potentiometer 
can see the change of the photocurrent voltage in the first voltmeter
and the fluctuations in the second voltmeter.
The first step of education is a qualitative description of the effect.
In a sensual qualitative level, even a person with very basic physics background can feel how 
the set-up works, which is described in the next section.

\section{The experimental set-up}
\label{sec:set-up}
\begin{figure*}[t]
\centering
\includegraphics[width=\textwidth]{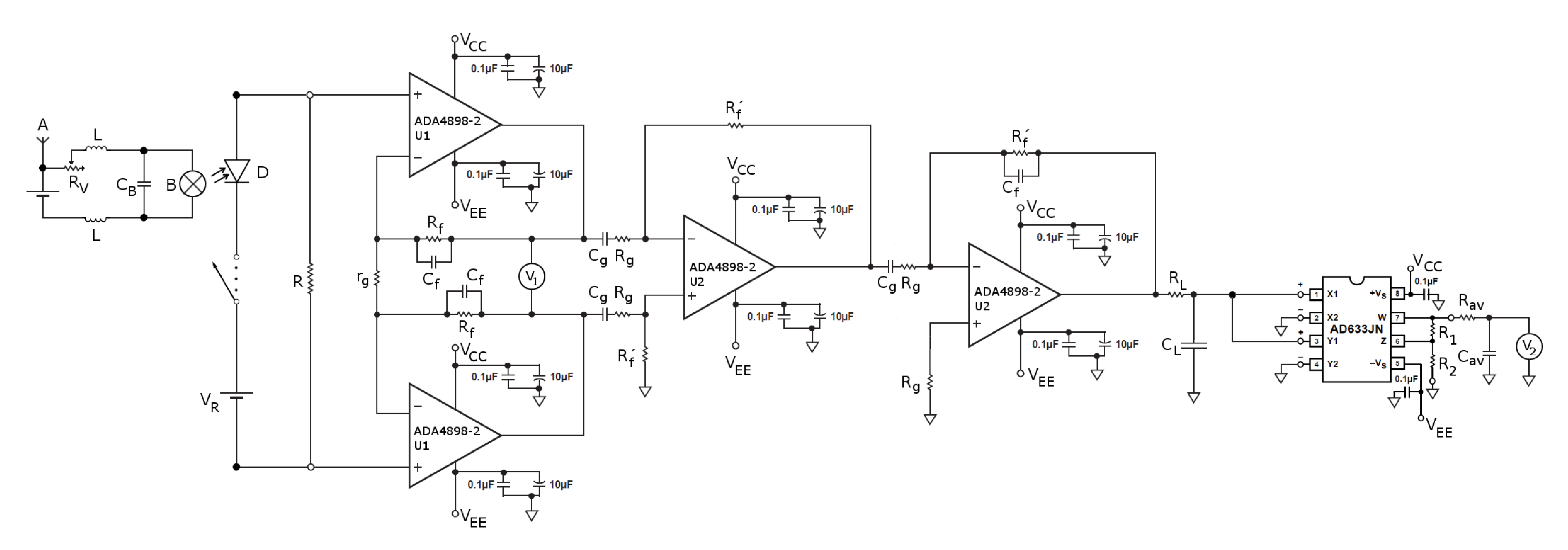}
\caption{Experimental circuit for $q_e$ measurement.
Light from the incandescent light bulb B shines on the photodiode D, which creates a photo current.
The photo current passes through the connected resistor $R$,
and creates a photo voltage $U(t)$
at the inputs of the instrumentation amplifier followed by an inverting amplifier.
The high impedance buffer of the instrumentation amplifier 
amplifies the time averaged photo voltage  
$U_1 = y_1 \left < U(t) \right >$ to be measured by multimeter V$_1$.
Alongside the DC photo voltage, Schottky noise from the photodiode is amplified
by the high impedance buffer too. 
The gain capacitors $C_\mathrm{g}$ 
stop the operational amplifiers offset and the voltage DC component 
of the amplified photo voltage $U_1$.
The amplified shot noise is further amplified by the difference and inverting amplifiers.
The amplified Schottky noise $U_\mathrm{amp}(t)$
is filtered by a low-pass filter, then is squared by a AD633 multiplier $U_{_\square}(t)$ and finally
it is averaged $U_2 = \left < U_{_\square}(t) \right >$ by a low-pass filter with a large time constant
$\tau_\mathrm{av} = R_\mathrm{av} C_\mathrm{av}$ to be measured by multimeter V$_2$.
The batteries for the light bulb and their connected cables pick up electromagnetic noise and
act like an effective antenna A.
The inductances $L$ are radial type choke coils with ferrite core CW68-102K and they switch off the circuit for higher frequencies, while
the capacitor $C_\mathrm{B}$ short circuits the remaining AC current component that 
has managed to pass through the inductances.
In such a manner, no AC current reaches the light bulb.
Sometimes only the capacitor is enough to switch off the antenna from the lamp.
The variable resistance $R_\mathrm{V}$ is used to change the current through the light bulb,
which in turn changes the luminosity of the latter, enabling measurements of
V$_1$ and V$_2$ for different values of photo voltage and Schottky noise.
The values of the used elements are given in Table~\ref{tbl:values}.}
\label{fig:set-up}
\end{figure*}

The circuit of the set-up is depicted in Fig.~\ref{fig:set-up}.
The light from a lamp creates a photo current through the photodiode, which passes through the connected resistor with resistance $R$.
The photo voltage $U(t)$ is amplified by an instrumentation amplifier followed by
an inverting amplifier.

The amplified voltage with total amplification $y$, which we will describe shortly after
\be
U_\mathrm{amp}(t)=yU(t)
\label{amp}
\ee
is applied to a multiplier giving output voltage $U_{_\square}(t)=U_\mathrm{amp}^2(t)/\tilde{U}$,
which is finally averaged by an averaging low-pass filter
with time constant $\tau_\mathrm{av}=R_\mathrm{av} C_\mathrm{av}$.
The voltage parameter $\tilde{U}$ is introduced later in this section.
The averaged voltage $U_2 = \left < U_{_\square}(t) \right >$
is measured by a voltmeter $V_2$,  a cheap commercial multimeter.
Another multimeter $V_1$ measures the time averaged photo voltage
\be
U_1 = y_1 \left < U (t) \right >,
\label{U1}
\ee
where $y_1$ is the amplification of the high-impedance buffer
\be
y_1=1+2R_\mathrm{f}/r_\mathrm{g},
\label{y1}
\ee
which contains 2 non-inverting amplifiers.
The gain capacitors right after the buffer stop the operational amplifiers 1/f noise and voltage offsets, as well as the time averaged DC photo voltage $U_1$.
Next there are a difference amplifier with amplification
$y_2=-R_\mathrm{f}^\prime/R_\mathrm{g}$
and an inverting amplifier with amplification
$y_3=-R_\mathrm{f}^\prime/R_\mathrm{g}$.
The amplifier has a total amplification
\be
y=y_1 y_2 y_3=\left(\frac{R_\mathrm{f}+r_\mathrm{g}/2}{r_\mathrm{g}/2}\right)
\left(\frac{R_\mathrm{f}^\prime}{R_\mathrm{g}}\right)^2.
\label{y}
\ee
It is possible to test each amplification $y_1, y_2, y_3$ by means of a simple voltage divider, using 1\% accuracy resistors.
After the voltage is reduced 100 times by the voltage divider, it is applied to the input of each amplification step (1 high impedance buffer, 2 difference amplifier and 3 inverting amplifier) and the voltage at the output should be equal to its initial value before the reduction, i.e. the input voltage should be recovered.
\begin{enumerate}
\item Input voltage of the first step should be applied between the (+) inputs of the buffer without removing the resistor $R$.
The recovered voltage should be measured with multimeter V$_1$.
The OpAmps participating in the high impedance buffer are placed in module U1 in Fig.~\ref{fig:set-up}.
The hidden secret of the whole circuit is the tiny distance between the (+) inputs of the dual operational amplifier.
The usage of a single OpAmp requires screening metallic boxes and BNC connecting cables.
\item For the difference amplifier, input voltage should be applied in the two points between $C_\mathrm{g}$ and $R_\mathrm{g}$ shown in Fig.~\ref{fig:set-up} before the left U2 OpAmp.
The recovered voltage should be measured between the output point and the common point of this OpAmp.
\item Input voltage of the inverting amplifier should be applied between the point between
$C_\mathrm{g}$ and $R_\mathrm{g}$ and the common point of the right U2 OpAmp.
The recovered voltage should be measured between the output voltage and common point of this last OpAmp.
\end{enumerate}
This simple method allows for a quick verification of the expected amplification using a DC input voltage source, for instance an ordinary battery.

In well equipped labs it is possible to use AC voltages to test two steps sequentially and even all three steps of the amplifier.
Only several centimeters of wire connected at the input of the amplifier creates ringing and parasitic capacitance between amplifier input and output.
Therefore extreme care is needed.
It is necessary to use triaxial cable and the external shielding to be connected to the common point of the amplifier.
Parallely to the resistor $R$ one $50~\Omega$ resistor should be connected to minimize reflected waves.
This triaxial cable should be connected to the circuit in a way to ensure the absence of ringing.
Only after that, one million time reduced AC voltage  in a frequency range between the lower cutoff frequency $f_\mathrm{g}$ and $f_\mathrm{_A}$ in Table~\ref{tbl:pars} from a signal generator
should be applied to the internal wire and the first shield of the triaxial cable.
In this way the reduced input signal will be amplified by the amplifier to recover the original signal from the signal generator.
We emphasize that recovery of signals after one million times amplification is a difficult task.
In case of external noise one can use lock-in amplifier for measuring of the amplified voltage.
Changing the frequency, one can investigate the frequency dependence of the transmission of the amplifier.
This procedure is standard for an electronics lab but it is far beyond educational measurement of a school lab.
That is why we recommend only the DC measurement of the different steps of the circuit.
Do not hesitate to write to the authors if you try repeating the experiment.

Usually specialists in electronics give the advice that care is needed to construct high gain 
and high-bandwidth amplifiers. 
This is absolutely true for the standard circuits used for the investigation of 
Schottky and Johnson noise and up to now screening metallic boxes and BNC connectors were indispensable.
However, for the instrumentation amplifier with dual OpAmps in the buffer 
parasitic capacitances are very small and ringing can come only by a large battery
(which is also an effective antenna for radio signals) of the photodiode.
To ensure circuit stability, care needs to be taken to minimize electrostatic coupling between
the photodiode battery (for instance, using small 3~V lithium battery) and 
million times amplified photovoltage fluctuations.
There are no screening boxes and no BNC connectors
in our set-up which is an advantage from a pedagogical point of view.  

After the amplification of $y$ times, the signal passes through a low-pass filter
with transmission coefficient 
\begin{equation}
y_{_\mathrm{LPF}}=\frac1{1+\mathrm{j} \omega\tau_{_\mathrm{LPF}}}, \quad
\tau_{_\mathrm{LPF}}=R_\mathrm{_L}C_\mathrm{_L},
\end{equation}
and bandwidth 
\begin{equation}
B_{_\mathrm{LPF}}=\int_0^\infty\frac{1}{1+\left(\omega\tau_{_\mathrm{LPF}}\right)^2}
\frac{\mathrm{d}\omega}{2\pi}=\frac1{4\tau_{_\mathrm{LPF}}}.
\end{equation}
This filter reduces the total noise and its purpose is to fix exactly the value of the pass bandwidth.
The values of $R_\mathrm{_L}$ and $C_\mathrm{_L}$ can be measured with 3 digits accuracy 
with a non-expensive LCR multimeter, while in university labs\cite{MIT:13}
commercial Butterworth filters have accuracy of 2\%. 
In such a way, the bandwidth $B_{_\mathrm{LPF}}$ of the simple low-pass filter can be evaluated with higher accuracy.

The standard operation of the multiplier with the voltages at its input terminals is described by the equation~\cite{AD633}
$W=(X_1-X_2)(Y_1-Y_2)/U_\mathrm{m}+Z$.
For our circuit both $X_2=Y_2=0$ and $X_1=Y_1=U_\mathrm{amp}(t)$.
In such a way, for the output voltage we have
\be
U_{_\square}(t) =W(t)=\frac{U_\mathrm{amp}^2(t)}{\tilde U}, \quad \tilde U \equiv U_\mathrm{m}\frac{R_1}{R_1+R_2},
\label{square}
\ee
where $U_\mathrm{m}$ is a constant of the multiplier.
Finally, the squared voltage $U_{_\square}(t)$ is averaged by the low pass filter
$U_2 = \left < U_{_\square}(t) \right >$
and measured by the voltmeter $V_2$.
Substitution of \Eqref{amp} in \Eqref{square} gives
\be
U_2 =\frac{(\delta U)^2}{U^*}, \quad
U^*\equiv \frac{\tilde U}{y^2}, \quad 
(\delta U)^2 \equiv \left < (U - \left < U\right >)^2 \right >.
\label{U2}
\ee
Here it is taken into account that the DC voltage $\left < U \right >$
is stopped by the first pair gain capacitors $C_\mathrm{g}$ immediately after the buffer.

According to the theory of Schottky noise, there is a linear dependence between the measured voltages,
which is described in the next section and the slope determines $q_e$.

In the experimental set-up
we use 2 dual low-noise operational amplifiers ADA4898-2,~\cite{ADA4898}
a cheap multiplier AD633,~\cite{AD633} a photodiode BPW34~\cite{BPW34}
(with small capacitance $C_\mathrm{ph}$)
powered by 3~V (CR1220) Li battery in reverse bias,
a T303 T3 3V 80~mA incandescent light bulb powered by 3~AA 1.5~V batteries
and 4 metal layer WIMA capacitors.~\cite{MKS2}

For the used operational amplifiers ADA4898-2 the 1/f noise dominates below the corner frequency $f_\llcorner$ on the order of 10~Hz.
Such low frequencies and DC offsets are stopped by the gain capacitors which together with gain resistors form lower cutoff frequency of the described amplifier 
$f_\mathrm{g}=1/2\pi R_\mathrm{g}C_\mathrm{g}= 159\;\mathrm{Hz}.$ 
In other words, the 1/f noise of the OpAmps is negligible for our circuit depicted in Fig.~\ref{fig:set-up}.
\begin{center}
\begin{table}[h]
\begin{tabular}{| c | r || c | r |}
		\hline
		Circuit element  & Value & Circuit element & Value \\ \tableline 
			$R$ & 200~$\Omega$ & $R_\mathrm{V}$ & 0-1~k$\Omega$ \\
			$C_\mathrm{B}$ & 10~$\mu$F & $R_1$ &  2~k$\Omega$ \\
			$C_\mathrm{_L}$ & 47~nF  & $R_2$ & 18~k$\Omega$\\ 
			$R_\mathrm{_L}$ & 510~$ \Omega$ & $R_\mathrm{av}$ & 1.5~M$\Omega$ \\
			$r_\mathrm{g}$ & 20~$\Omega$ &   $C_\mathrm{av}$ & 10~$\mu$F \\
			$R_\mathrm{f}$ &  1~k$\Omega$ &  $V_\mathrm{CC}$ & +9~V\\
			$C_\mathrm{f}$ &  10~pF &  $V_\mathrm{EE}$ & -9~V \\ 
			$C_\mathrm{g}$ & 10~$\mu$F & $U_\mathrm{m}$ & 10~V\\
			$R_\mathrm{g}$ &  100~$\Omega$ & $L$ & 1~mH \\ 
			$R_\mathrm{f}^\prime$ & 10~k$\Omega$ & $C_\mathrm{ph}$ & $\le 40~$pF \\
\tableline
\end{tabular}
	\caption{Table of the numerical values of the circuit elements from Fig.~\ref{fig:set-up}.}
	\label{tbl:values}
\end{table}
\end{center}
\begin{center}
\begin{table}[h]
\begin{tabular}{| c | r || c | r |}
		\hline 
		Parameter  & Value & Parameter & Value  \\ \tableline
	   		$\tau=RC_\mathrm{ph}$  & $\le 8\;\mathrm{ns}$ & $y_1$ & 101 \\
			$f_0$ & 65~MHz~\cite{ADA4898} & $y_2$ & -100  \\
			$\tau_0$ & 2.45~ns & $y_3$ & -100 \\       
			$\tau_\mathrm{_A}$ & 247~ns &  $y$ & $1.01 \times 10^6$ \\
			$f_\mathrm{_A}$ & 669~kHz & $U_\mathrm{m}$ &  10~V~\cite{AD633} \\
			$\tau_{_\mathrm{LPF}}$ & 23.97~$\mu$s &  $\tilde U$ & 1~V  \\
			$\tau_\mathrm{f} $ & 10~ns & $U^*$ & 891~fV \\
			$\tau_\mathrm{g}$ & 1~ms & $G^2$ & 103~dB\cite{ADA4898}  \\
			$f_\mathrm{g}$ & 159~Hz & $G$ & 10$^5$ \\
			$\tau_\mathrm{f}^\prime$ & 100~ns &  Input voltage noise & 0.9~nV/$\sqrt{\mathrm{Hz}}$  \cite{ADA4898}\\
                $\tau_\mathrm{av}$ & 15~s & CMRR (typical )& -126~dB~\cite{ADA4898} \\
\tableline
\end{tabular}
	\caption{Table of the calculated parameters necessary for the analysis of the circuit
	(time constants, frequencies, voltages and linear amplification).
$f_0$ is the -3dB bandwidth of the ADA4898, $\tau_0\equiv 1/2\pi f_0,$
$f_{_\mathrm A} \equiv f_0/y_1,$ $\tau_{_\mathrm A}=\tau_0 y_1,$
$\tau_{_\mathrm{LPF}}=R_\mathrm{_L}C_\mathrm{_L},$ $\tau_\mathrm{f}=R_{\mathrm f}C_{\mathrm f},$
$\tau_\mathrm{f}^\prime=R_{\mathrm f}^\prime C_{\mathrm f}^\prime,$
$\tau_\mathrm{g}=R_{\mathrm g}C_{\mathrm g},$
$\tau_\mathrm{av}=R_{\mathrm {av}} C_{\mathrm {av}},$
and \\
$\tau_0\ll \tau_\mathrm{f}, \tau_\mathrm{f}^\prime, \tau_{_\mathrm A}
\ll \tau_{_\mathrm{LPF}}\ll \tau_{\mathrm g}\ll \tau_\mathrm{av}$.}
	\label{tbl:pars}
\end{table}
\end{center}
The other elements are ordinary 1\% accuracy resistors and the rest of the capacitors
are within 20\% accuracy.
The exact value of $C_\mathrm{_L}$ is essential and therefore
has to be measured with high accuracy.
The values of the used elements are given in Table~\ref{tbl:values}.


\section{Recalling shot-noise theory}
\label{sec:theory}
When the current through the photodiode consists of separate
$\delta$-shaped electron impulses at arbitrary times $t_\mathrm{i}$
\be
I(t)=\sum_\mathrm{i} q_e \delta(t-t_\mathrm{i}),
\ee
then between the average current
\be
\left < I \right > = \lim_{T\rightarrow\infty} \int_0^T I(t)\, \frac{\mathrm{d}t}T
\ee
and the low frequency spectral density $(I^2)_f$  current, which
parameterizes the current dispersion
\be
(\delta I)^2 = \left < (I-\left< I \right >)^2 \right > =
\int_0^\infty (I^2)_f \frac{\mathrm{d} \omega}{2 \pi}
\ee
the Schottky law~\cite{Schottky:18} is in effect
\be
(I^2)_f = 2 q_e \left < I \right >.
\ee
The theory of shot noise is given in Appendix~\ref{shot-noise} of the current work.

In such a way for the spectral density of the voltage we obtain
\be
(U^2)_f= 2 q_e \left < I \right > R^2+ 4k_\mathrm{_B}T R+
e_\mathrm{_N}^2+R^2 i_\mathrm{_N}^2,
\ee
where the second term describes the thermal noise of the resistor $ 4k_\mathrm{_B}T R$,
the third $e_\mathrm{_N}^2$ gives the contribution of the voltage noise of the operational amplifier
and the last term expresses the current noise $i_\mathrm{_N}^2$ of the operational amplifier.
The detailed theory of Johnson noise will be given in another paper.

Taking into account the bandwidth $B_{_\mathrm{LPF}}$ of the low pass filter 
we obtain for the voltage dispersion
\be
(\delta U)^2 = \left < (U - \left < U\right >)^2 \right > =
\int_0^\infty (U^2)_f \frac{ \mathrm{d} \omega}{2 \pi}
=\frac{q_e}{2C_\mathrm{_L}} \frac{R}{R_\mathrm{_L}} \left< U\right>,
\label{vdisp}
\ee
where $\left < U\right > = R \left < I\right >$ is the time averaged photo voltage of the resistor.
In this formula for the Schottky noise we express $\left< U \right>$ and $\delta U$ 
by the experimentally measurable $U_1$ from \Eqref{U1} and $U_2$ from \Eqref{U2}
and obtain 
\be
U_2 =\alpha U_1 + v,\quad \alpha \equiv \frac{\Delta U_2}{\Delta U_1}=
\frac{q_e R}{2 R_\mathrm{_L} C_\mathrm{_L} y_1U^*},
\quad v=\mathrm{const}.
\ee
In such a way the electron charge $q_e$ can be expressed by the dimensionless slope $\alpha$ 
of the linear regression
\be
q_e=2 \alpha y_1 \frac{R_\mathrm{_L}}{R} C_\mathrm{_L}U^* .
\ee
Substituting here $y_1$ from \Eqref{y1} and
$U^*$ from \Eqref{U2} and $\tilde U$ from \Eqref{square}
finally the electron charge is expressed by the experimentally measurable values
\be
q_e
=\frac{2 r_\mathrm{g}}{2R_\mathrm{f}\!+\!r_\mathrm{g}}
\!\left(\frac{R_\mathrm{g}}{R_\mathrm{f}^\prime}\right)^{\!\! 4}\!
\frac{R_\mathrm{_L}C_\mathrm{_L} U_\mathrm{m} }{R} 
\frac{R_1}{R_1\!+\!R_2}
\frac{R_\mathrm{V}\!+\!R_\mathrm{av}}{\mathcal{Z} R_\mathrm{V}}
\frac{\Delta U_2}{\Delta U_1}\!.
\label{qe}
\ee
Here the last multiplier $R_\mathrm{V}/(R_\mathrm{V}+R_\mathrm{av})$ describes the 
voltage divider created by the finite internal resistance of the multimeter $R_\mathrm{V}$
(1~M$\Omega$ in our case) and the averaging low pass filter resistance $R_\mathrm{av}$.
The calculation of the bandwidth factor $\mathcal{Z}=1-\epsilon \approx 1$ taking into account the operational amplifiers' time constant $\tau_0$ is calculated in Ref.~\cite{MYV}. 
For the elements values used in the described experiment here, the correction is $\epsilon=6.74$\% and it is included in the obtained value for the electron charge.
This correction takes into account the finite value of the crossover frequency $f_0$ of the operational amplifier.
The inclusion of this correction is necessary only if we wish to reach 1\% accuracy of the measurement of $q_e$.
In the next section we will describe how this idea for the determination of $q_e$ is realized.

\section{Experiment}
\label{sec:experiment}
\begin{figure}[h]
\centering
\includegraphics[scale=0.6]{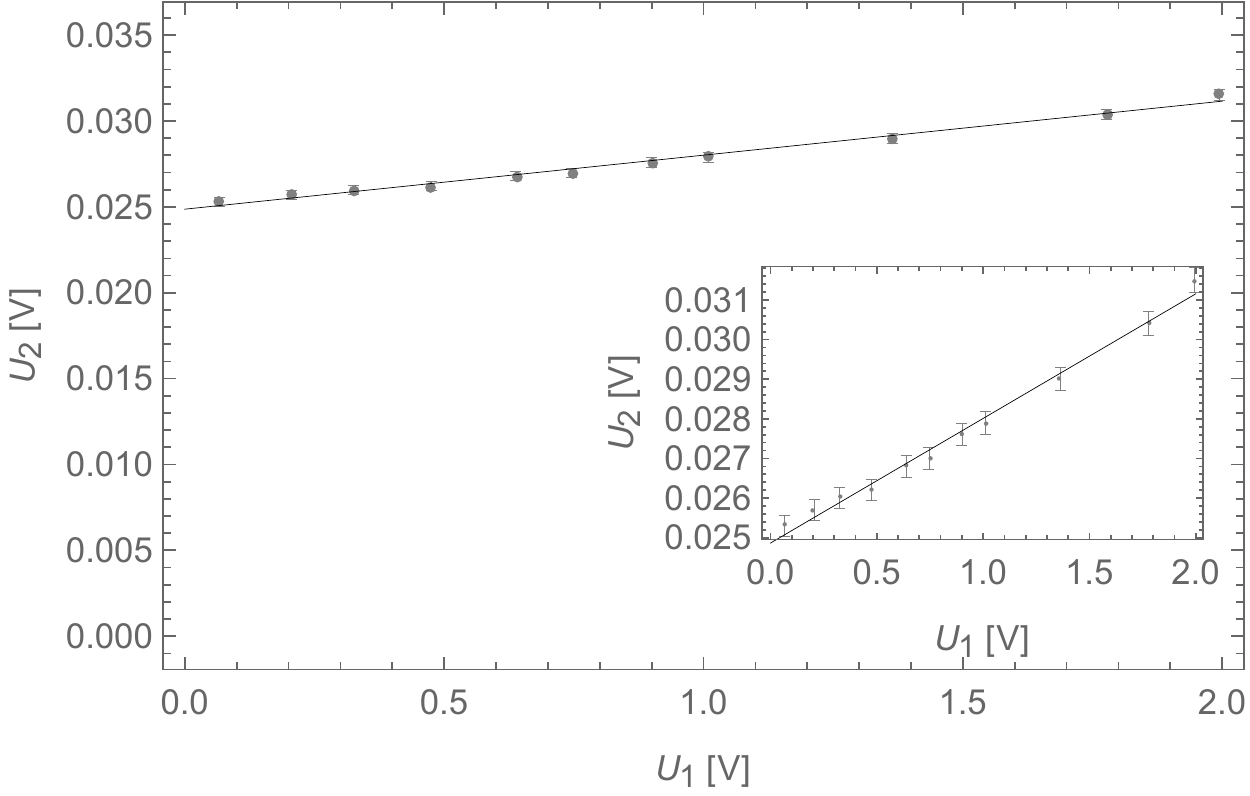}
\caption{
$U_2$ versus $U_1$ plot.
Linear regression from the electron charge $q_e$
measurement data given in Table~\ref{tbl:exp}.
According to \Eqref{qe} the slope $\alpha$ of the linear regression $U_2=\alpha U_1+v$
determines the electron charge $q_e$.
The experimental data procession gives $\alpha=3.145\times 10^{-3}$ with correlation coefficient 
$\rho=0.995$.
The abscissa voltage $U_1\propto \left<U\right>$
is proportional to the time averaged photo current $\left<I\right>$
while the ordinate voltage $U_2\propto (\delta U)^2$
is proportional to the dispersion of the shot noise.
The calculated electron charge $q_e$ with the obtained slope $\alpha$
is $(1.811\pm 13\% )\times 10^{-19}$~C, which is approximately 13\% larger than
the real value $1.602 \times 10^{-19}$~C. 
The constant $v \approx 0.025$~V of the linear regression is the voltage offset
that includes the thermal noise of $R$ and the voltage noise of the first couple
operational amplifiers mainly.
The large value of $v$ (approximately 5 times the difference between the minimal and maximal
values of $U_2$) shows how crucial the usage of low-noise operational amplifiers is. 
The inset shows the segment of the linear dependence 
with 1\% error bars of $U_2$ corresponding to statistical evaluation of many similar experiments 
and deviation of $\rho$ from one.}
\label{fig:regr}
\end{figure}

For fixed light intensity the averaging time is on the order of one minute.
The voltage proportional to the dispersion of the shot noise $U_2$ 
has more significant fluctuations than amplified DC voltage $U_1.$
The measurements made within an hour are presented 
in Table~\ref{tbl:exp}
and graphically depicted in Fig.~\ref{fig:regr},
the obtained value for the electron charge $q_e$ from these measurements is
$(1.811\pm 13\% )\times 10^{-19}$~C, which is 13\% higher than its true value.
\begin{table}[h]
\centering
\begin{tabular}{| l | l |}
		\hline
		&  \\ [-1em]
		$U_1$ [V] & $U_2$  [V] \\ \tableline
			 &  \\ [-1em]
			0.067 & 0.0253 \\
			0.205 & 0.0257 \\
			0.325 & 0.0260 \\
			0.476 & 0.0262  \\
			0.638 & 0.0268 \\ 
			0.745 & 0.0270\\
			0.900 & 0.0276\\
			1.010 & 0.0279 \\
			1.364 & 0.0290 \\
			1.777 & 0.0304\\
			1.994 & 0.0315\\
\tableline
\end{tabular}
\caption{Data of the voltages given by voltmeters V$_1$ and  V$_2$ 
from Fig.~\ref{fig:set-up}
for the electron charge $q_e$ measurement.
The the graphical representation and linear regression $\Delta U_1/\Delta U_2$ 
are shown in Fig.~\ref{fig:regr}.}
\label{tbl:exp}
\end{table}

The accuracy can be increased by increasing the time averaging constant  $\tau_\mathrm{av}$,
screening the set-up from the external noises, different battery supplies for different operational amplifiers, taking non-ideal effects of the operational amplifier described by the master equation for the OpAmps 
\be
\left (G^{-1}+\tau_0 \frac{\mathrm{d}}{\mathrm{d}t} \right )U_\mathrm{output}=U_+-U_-,
\ee
giving the relation between the inputs $U_+$ and $U_-$
and the output $U_\mathrm{output}$ of the OpAmp;
for the numerical values of the parameters see Table~\ref{tbl:pars}), 
precise measurement of the values of all components etc. 
The master equation of the OpAmps that apparently has never been published before
(the authors will very much appreciate receiving an appropriate reference with the equation present in it) is described in our recent eprint~\cite{Master}
and the results of its application can be found in many technical applications.
See for example the formulae for the frequency dependence of the inverting amplifier and non-inverting 
amplifiers in the specification of ADA4817.\cite{ADA4817}

In such a way, we can achieve the accuracy of the good university students labs
\cite{Earl:66,Livesey:73,Kittel:78,Vetterling:79,Ericson:88,Kraftmakher:95,Spiegel:95,
Kraftmakher:05,Cristofolini:03,Rodriguez:10,Wash:12,MIT:13,Dessau:16}
using standard electronics
with low noise pre-amplifiers and Butterworth filters with 2\% accuracy.
But the purpose of the present work is in a different direction. 

The amplification of our set-up is on the order of that used by the Habicht~\cite{Habicht:10} brothers trying to measure the Boltzmann constant  $k_\mathrm{_B}$ as it was suggested by Einstein~\cite{Einstein:07}.
For the electronic amplifier the problem of the floating of the zero was solved by the gain capacitors. 
For our set-up the calibration of the amplifier should be checked by determination of the 
Boltzmann constant. 
Having removed the photodiode we have to change the resistor $R$ with several values to perform a linear regression in the plot 
$U_2$ versus $R.$  The slope of the linear regression then gives $k_\mathrm{_B}$.

\section{Discussion and Conclusions}
\label{sec:conclusion}
The goal of our work is not a competition with the best students laboratories but to distribute the set-up in all universities and to consider even high school labs.
Our approach does not require specialized (in that sense expensive) laboratory equipment.
A soldering station and two multimeters can be found in every high school physics lab.
No such opportunity to built cheap and widely accessible experimental set-up is offered by any of the published works on Johnson and Schottky noise.

Working in extremely noisy environment, for instance in the presence of luminescent light, it is better to use electromagnetic screening.
A small fridge is absolutely enough and moreover, cooling significantly improves the work of the photodiode.
Our set-up works very well in temperatures lower than 15~$^\circ$C.
And vice versa, a hair dryer can destabilise the work of the set-up and give spurious and different values for the electron charge $q_e$.

It is extremely stimulating to measure a fundamental constant with do-it-yourself set-up.
But we can consider the next step.
Industrial production of the present set-up using PCB and automatized  soldering of elements can lead to a price which is comparable to that of a scientific calculator acceptable for every high school around the world. 
The electron charge $q_e$ is included in the high school education programs all around the globe.
It is time to include its measurement as a standard lab practice.
The electron and electronics change the life of the people and it is time all people using electronic devices to remember how they measured $q_e$ at teen age. 
Last but not least, this experiment is fun.


\section{Notes added at proof}
\label{sec:proof}

In Sec.~\ref{sec:set-up} we have derived the total amplification $y$ and the separate amplifications of the high-impedance buffer $y_1$, the difference amplifier $y_2$ and the inverting amplifier $y_3$ neglecting the frequency response of the operational amplifiers.
For high-school or even freshman university students this is quite acceptable.
But for complete understanding of the set-up as an engineering device, a frequency dependent analysis is necessary.
This analysis has already been thoroughly done and the correction factor $\mathcal{Z}$ is also introduced and explained~\cite{MYV}.
This correction factor depends on the capacity of the feedback $C_\mathrm{f}$ and gain $C_\mathrm{g}$ capacitors.
In the ideal case considered in this work $C_\mathrm{f} = 0$, $C_\mathrm{g} = \infty$ and $f_0=\infty$ and the total amplification $y$ is given by \Eqref{y}.
Considering the non-ideal case (the frequency dependence), the total amplification $y(\omega)$ is frequency dependent and is given by the tedious formula in Eq.~(A25) of Ref.~\onlinecite{MYV}.
And the introduced here correction factor $\mathcal{Z}$ is calculated numerically by Eq.~(A33) of Ref.~\onlinecite{MYV}.

In addition, several high-school oriented problems on Ohm's law are given in Subsec.~\ref{subs:OL}.

A schematic representation of the PCB SOIC-to-DIP converter and experimental set-up board are given in Subsec.~\ref{subs:GF}. \\


\acknowledgments{}

The authors are grateful to Vasil Yordanov for the friendship and his contribution 
at the early stages of the present research~\cite{MYV} (for an extended bibliography and history of the problem see this unabridged version), to Alexander Petkov for making the first measurements and critical reading of the manuscript, to Nikolay Zografov for introducing order in the lab, to Andreana Andreeva for animation of the spirit in the lab and to Petar Todorov for the interest and assistance in the current research.

One of the authors TMM is thankful to Genka Dinekova and Pancho Cholakov for recommending
ADA4898 \cite{ADA4898} in the beginning of the present work and for a valuable bit of advice related to the integrated circuits tested in the development of our set-up.
University of Sofia St. Clement of Ohrid received many free of charge operational amplifiers from Analog Devices, which is actually an important help to our university research and education.

We appreciate the fruitful discussions with Marina Poposka and Riste Popeski-Dimovski.

The present measurement of the electron charge $q_e$ is one of the series of experimental set-ups
created by the newly founded educational Laboratory for measurements of fundamental constants.
Here we wish to mention the experimental set-ups for measuring of the Planck constant $\hbar$~\cite{Planck}, the velocity of light $c$~\cite{Light} and most recently the Boltzmann constant $k_\mathrm{_B}$~\cite{Boltzmann}, which are in the educational programme of the laboratory.



\appendix


\section{Pedagogical rederivation of Schottky formula for shot noise}
\label{shot-noise}

\begin{figure}[h]
\begin{center}
\includegraphics[scale=0.3]{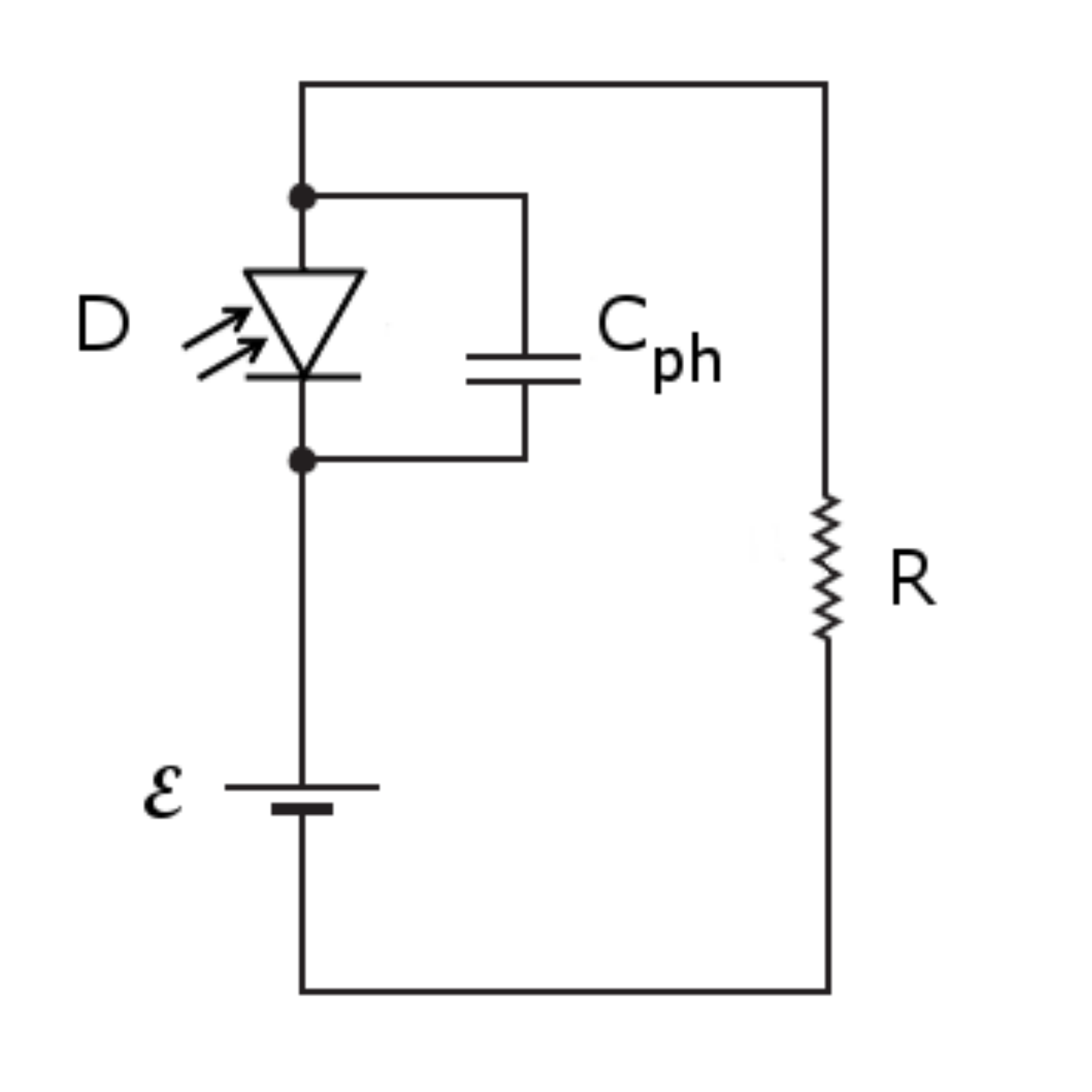}
\caption{Schottky shot noise produced by the discrete
photo-current through the photo-diode D.}
\label{fig:schotky_noise}
\end{center}
\end{figure}

Let us analyze the current impulse of a single electron through a resistor 
\begin{align}
&J(t)=q_e\delta_\tau(t-t_1), \\
& \delta_\tau(t)\equiv\frac{\theta(t)}{\tau}\exp(-t/\tau), 
\quad \tau=R C_\mathrm{ph} \nonumber
\end{align}
created by the quantum transition of one electron with charge $q_e$ in the photo-diode D with capacity
$C_\mathrm{ph}$; see Fig.~\ref{fig:schotky_noise}.
One can imagine a real photodiode as parallely connected ideal photodiode and a small capacitor with capacitance $C_\mathrm{ph}$.
This capacitor together with the shunting resistor $R$ creates a circuit with time constant $\tau$ and corresponding characteristic frequency
\be
f_\mathrm{ph}=\frac{1}{2 \pi \tau} = \frac1{2\pi R C_\mathrm{ph}}.
\ee
Supposing that after some long time $M$ (say one hour) the current is repeated $I(t+M)=I(t)$,
we can use Fourier representation
\begin{align}
&J(t)=\left<J\right> +\sum_{\omega\ne0} J_\omega\exp(-\mathrm{i}\omega t), \\
&\left<J\right>=\int_0^M J(t)\frac{\mathrm{d}t}{M}, \quad
\omega=\frac{2\pi}{M}n, \quad n=0,\pm1,\pm2,\pm3,\dots, \nonumber
\end{align}
where in our case
\begin{align}
&J_\omega=\int_0^M J(t) \exp(\mathrm{i}\omega t) \frac{\mathrm{d}t}{M}
=\frac{q_e}{M}\frac{\exp(\mathrm{i}\omega t_1)}{1-\mathrm{i}\omega\tau},\\
& \label{Y_ph}
\left|\Upsilon_\mathrm{ph}(\omega)\right|^2\equiv
\frac{M^2}{q_e^2} \left|J_\omega\right|^2=\frac1{1+(\omega\tau)^2}=\frac1{1+(f/f_\mathrm{ph})^2}. \nn
\end{align}
Substituting in $M \rightarrow \infty$ the Parseval's theorem 
\begin{align}
\left<J^2\right>=
\int_0^M J^2(t)\frac{\mathrm{d}t}{M}
=\left<J\right>^2+\sum_{\omega\ne0} \left|J_\omega\right|^2,
\end{align}
substituting summation with integration
\begin{align}
\sum_{\omega} \approx M\int\frac{\mathrm{d}\omega}{2\pi}
\end{align}
we express the fluctuation $\delta J\equiv\sqrt{\left<(J-\left<J\right>)^2\right>}$ 
by the spectral density $(J^2)_{\!f}\equiv 2 \left|J_\omega\right|^2$
\begin{align}
\label{density}
(\delta J)^2=\left<(J-\left<J\right>)^2\right>=M\int_0^\infty (J^2)_{\!f} \frac{\mathrm{d}\omega}{2\pi}.
\end{align}
For real $J(t)$ the spectral density is even $(J^2)_{\!f}(\omega)=(J^2)_{\!f} (-\omega)$.

If the shot noise in time interval $M$ is created by $N$ incoherent and independent quantum transitions
\begin{align}
&I(t)=\sum_{a=1}^N q_e\delta_\tau(t-t_a),\quad 
0<t_1<t_2<\dots<t_N<M,\\
&\left<I\right>=\frac{q_e N}{M},
\quad N=\frac{\left<I\right> M}{q_e}.
\end{align}
For the fluctuation of the current, its dispersion and spectral density we have just to multiply 
by the number of the different quantum transitions
\begin{align}
(I^2)_{\!f}=N(J^2)_{\!f}.
\end{align}
We suppose that the moments $t_a$ of the different quantum transitions are completely independent.
Then substituting \Eqref{Y_ph} in \Eqref{density}
we arrive to the Schottky formula for the spectral density of the current noise
\begin{align}
&(\delta I)^2=\left<(I-\left<I\right>)^2\right>=\int_0^\infty (I^2)_{\!f} \frac{\mathrm{d}\omega}{2\pi},\\
&(I^2)_{\!f} =2q_e\left<I\right>\left|\Upsilon_\mathrm{ph}(\omega)\right|^2,\\
&\left|\Upsilon_\mathrm{ph}(\omega)\right|^2\approx 1, \text{ for }  f\ll f_\mathrm{ph}.
\end{align}
As a rule the capacity of the photo-diode $C_\mathrm{ph}$ is negligible 
and the corresponding frequency $f_\mathrm{ph}$ is much higher than the working frequencies 
of the rest electronics. 
In this case, we have the well-known formula for the frequency independent
spectral density of the shot noise 
\begin{align}
\label{Schottky_noise}
(I^2)_{\!f} \approx 2q_e\left<I\right>,\quad \text{ for } f\ll f_\mathrm{ph}.
\end{align}
For the spectral density of the voltage, we analogously have 
the spectral density of the voltage for the Schotky shot noise is
\begin{equation}
(U^2)_{\!f}
= 2R^2 q_\mathrm{e} \left\langle I \right\rangle.
\end{equation}
This white noise approximation
corresponds for $\tau \rightarrow 0$ 
to $\delta$-function approximation in the time representation
\begin{align}
\delta_\tau(t)=\frac{\theta(t)}{\tau} \exp(-t/\tau)\approx \delta(t),
\quad \delta(t) \equiv \frac{\mathrm{d}\theta(t)}{\mathrm{d}t}.
\end{align}
Even Oliver Heaviside knew how the derivative of his $\theta$-function looks like~\cite{Jemmer:67},
but many years later defending from the mathematicians Dirac said 
``Every electrical engineer knows the term impulse, and this
function just represents it in a mathematical form"
Ref.~78 of the book by Jemmer~\cite{Jemmer:67}.
In our case the $\delta$-function approximation of the uncorrelated current impulses of
the different photo-electrons is applicable if the corresponding time constant
is much smaller than all other time constants of the circuit 
$\tau=R C_\mathrm{ph}\ll \tau_{_\mathrm{LPF}} \ll \tau_\mathrm{A}.$
No doubts frequency dependence of $\Upsilon_\mathrm{ph}$ from Eq.~\ref{Y_ph} and 
Eq.~\ref{Schottky_noise} can be easily programmed but it will be ignorantly taking into account for
the accuracy of a student laboratory.

\begin{figure}[h]
\includegraphics[scale=0.2]{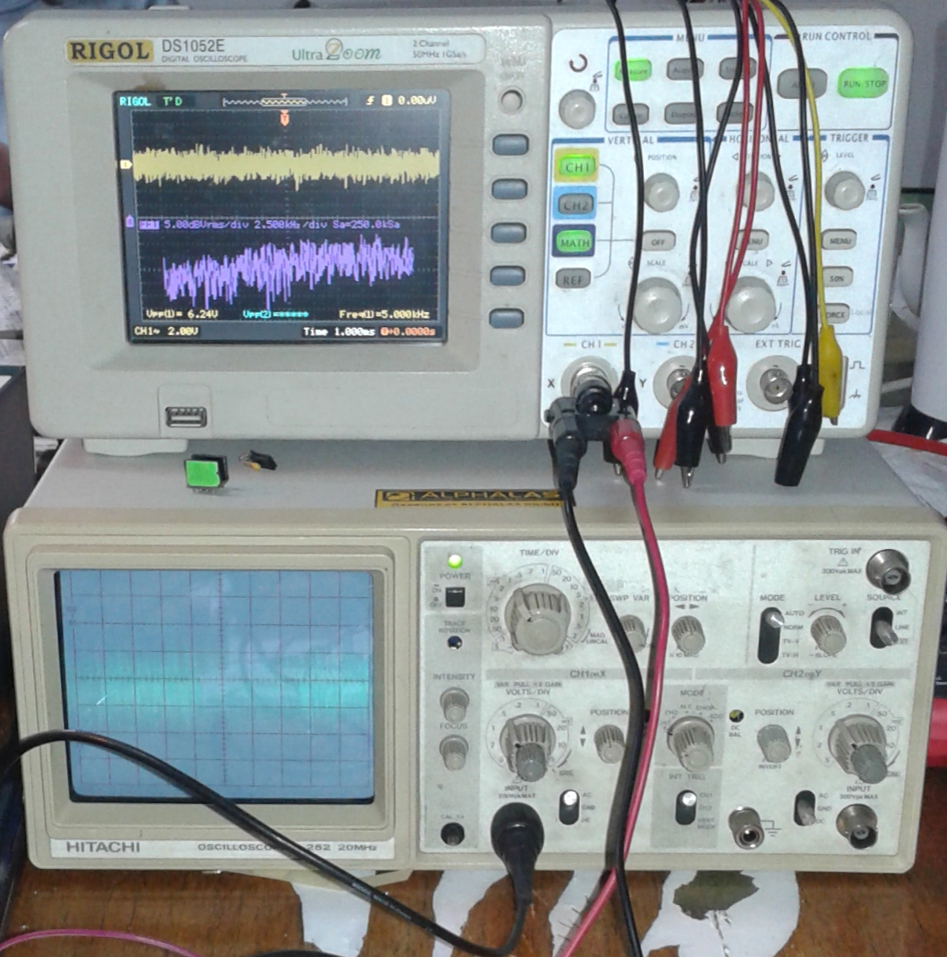}
\caption{Typical white noise created by photo-current.
In the photo one can see the time dependent current $I(t)$ observed by two oscilloscopes.
In the lower part of the upper oscilloscope one can see
the spectral density of the current $(I^2)_{\!f}(\omega)\approx \text{const}$
which is white enough.
The inflections describe the work of the oscilloscope and its fast Fourier transformation. 
}
\label{fig:noise}
\end{figure}

Returning back to the work of our set-up 
after one million times amplification the photo-current
transformed into voltage can be seen in the photo of two oscilloscopes
Fig.~\ref{fig:noise}.

Here we will allow ourselves a lateral speculation on the current noise of OpAmps.
For OpAmps the input bias current $I_\mathrm{B}$ is always small and obviously
related to incoherent quantum transitions through an insulating barrier.

For example, tunneling from the gate to source-drain channel in a field effect transistor through the narrow insulating barrier.
It is not always specified on data sheets, but the spectral density of the input current noise may be calculated in cases like simple BJT or JFETs, where all the bias current flows in the input junction, because in these
cases it is simply the Schottky noise of the bias current~\cite{MT047}
$i_\mathrm{n}^2 \approx 2q_\mathrm{e} I_\mathrm{B}$.

According to our interpretation, for electro-meter operational amplifiers current noise is just spectral density of the noise 
$i_\mathrm{n}^2=(I^2)_{\!f}$,
the average current is the input bias current 
$\left<I\right>=i_\mathrm{B}$
and these parameters are related by the Schottky formula.

For a verification with logarithmic accuracy we can use even data sheets and plot a linear regression
$\ln i_\mathrm{n}$ versus $\ln  I_\mathrm{B}$
\begin{equation}
2\ln i_\mathrm{n}\approx \ln  I_\mathrm{B}+\ln(2q_e).
\end{equation}

If we use electro-meter OpAmp AD549~\cite{AD549,low-ib} with extremely low bias current of $I_B=200$~fA
the whole bias current is related with incoherent processes of quantum tunneling and we can use thermal evaporation of the charge carriers from the gate as a perfect source of shot noise for determination of $q_e$ without photo-current using only integrated circuits.
It deserves to check with maximal accuracy whether the temperature dependence of 
electrometer operational amplifier is canceled in a temperature $T$ independent ratio
\begin{equation}
q_e=\frac{i_\mathrm{n}^2(T)}{2 I_\mathrm{B}(T)}.
\end{equation}
In this case Schottky formula will be checked by incoherent quantum transition from the gate to the source-drain channel using not a set-up but only one commercial integrated circuit to measure the electron charge.

In conclusion, the electron charge can be determined by measuring the input bias current of a good electrometer operational amplifier.
We suggest a new experiment to be performed with this method.
After all, the electron charge is the beginning of the electronics.

\section{Guide for building of the set-up}

The used double ADA4898-2 operational amplifier is available only in small $4\times4\mbox{ mm}$ SMD
(Surface Mount Device) package.
It is strongly believed that soldering of SMD components requires specific equipment and personal skills. 
Here authors have to add some observations in this point:
in the video abstract of this article is filmed how the doyen of the authors,
a professor in theoretical and mathematical physics can solder the ADA4898-2 operational amplifiers on 
a SOIC to DIP printed circuit board (PCB) adapter.
All other coauthors are much better in soldering with 10 decibels larger speed. 
One full professor in theoretical physics gives a good approximation for a motivated but awkward high-school student.
For training it is possible to start with some cheep op-amp having the same pin orientations, for example TL072.

\begin{figure}[h]
\includegraphics[scale=0.1]{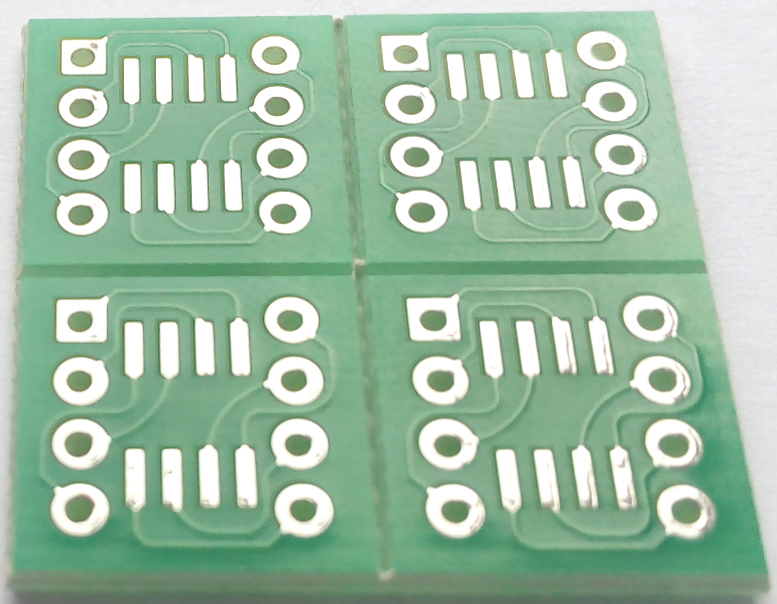}
\caption{Photo of SOIC Small Out-line Integrated Circuit)-to-DIP (Dual In-line package) adapter. 
We use it to connect ADA4898-2 operational amplifier to the main circuit.}
\label{fig:soic-to-dip}
\end{figure}

\begin{figure}[h]
\includegraphics[scale=0.1]{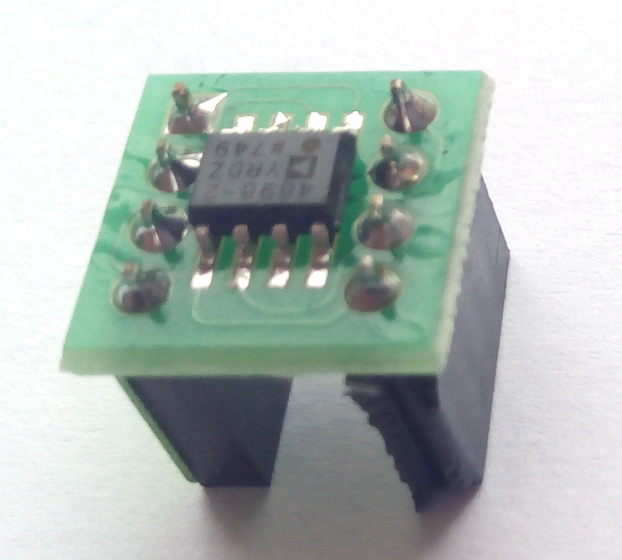}
\caption{Photo of the removable to the circuit adapter with ADA4898-2 on top.}
\label{fig:4898}
\end{figure}

\begin{figure}[h]
\includegraphics[scale=0.06]{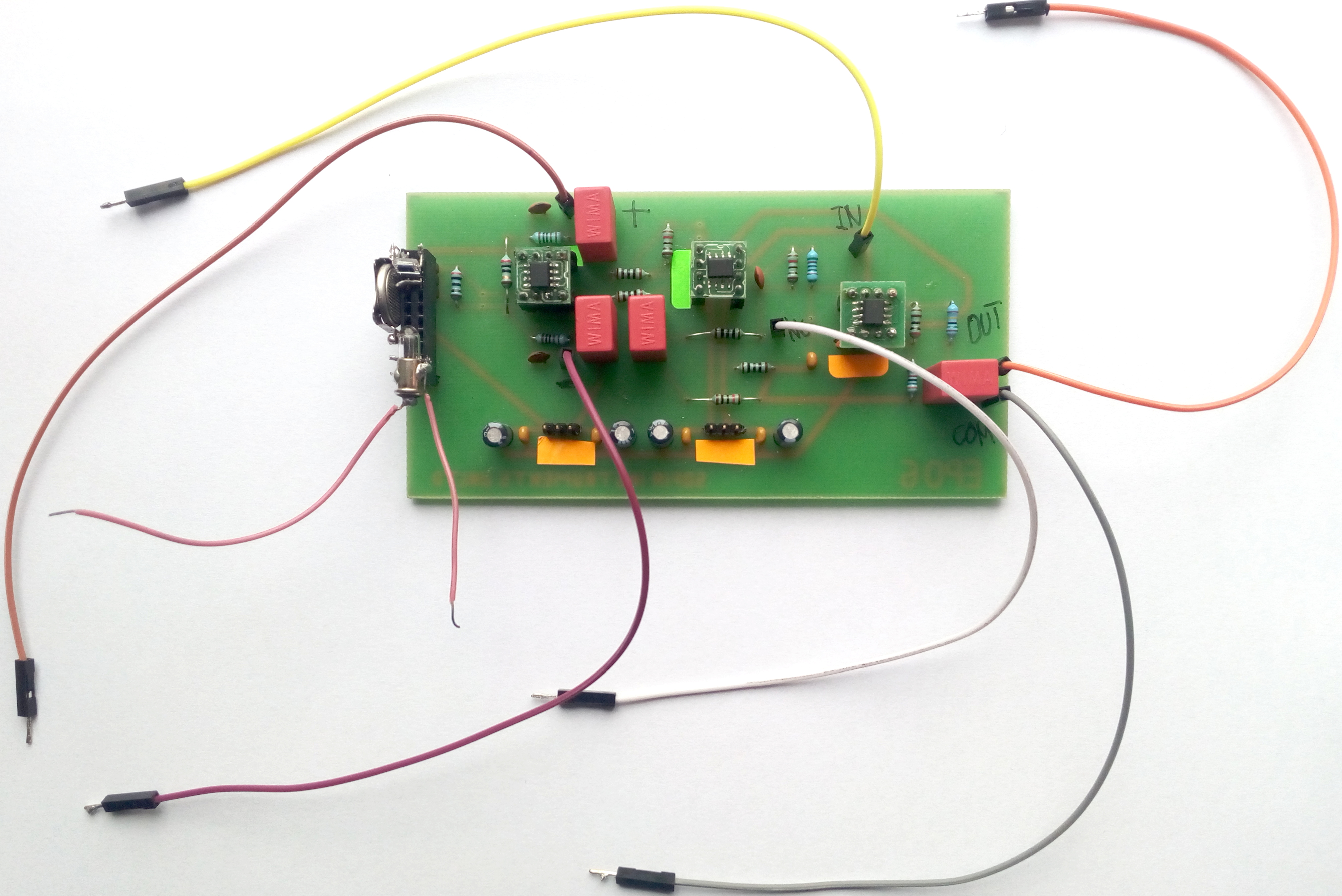}
\caption{Photo of the set-up. The electronic scheme of the circuit is depicted in Fig.~\ref{fig:set-up} and the board layout is shown in Fig.~\ref{fig:board}.}
\label{fig:photo_set-up}
\end{figure}

The main circuit can be prepared by a prototype PCB or in a specially designed PCB if we need to prepare more than 10 set-ups. For encouraging of reproducing of the set-up we are giving 
photos for the SOIC-to-DIP adapter Fig.~\ref{fig:soic-to-dip}, a SMD OpAmp mounted on a removable 
pedestal to the main circuit Fig.~\ref{fig:4898} and the main set-up Fig.~\ref{fig:photo_set-up}.
The Geber files of the PCBs are available as a supplementary material to the article.



\section{Supplementary material}

\subsection{Several simple problems on Ohm's law}
\label{subs:OL}

The frequency independent analysis of the experimental set-up described here is simple enough to be performed by high school students.
A different experiment with this amplifier has already been given to high school students~\cite{EPO5} and in this way, the amplifier is approbated for high school education.
The solutions of several simple problems on Ohm's law in section ``Several simple tasks on Ohm’s law'' of \onlinecite{EPO5} gives the theory of operation of the experimental set-up given in Sec.~\ref{sec:set-up}.
In this way, a dedicated high-school student can easily derive the theory of the work of the set-up.

\subsection{Schematic representations of the PCB boards}
\label{subs:GF}

In this subsection of the appendix, we include the schematic representations of the SOIC-to-DIP converter photographed in Fig.~\ref{fig:soic-to-dip} and the experimental set-up PCB board photographed in Fig.~\ref{fig:photo_set-up}.

\begin{figure}[h]
\centering
\includegraphics[scale=0.3]{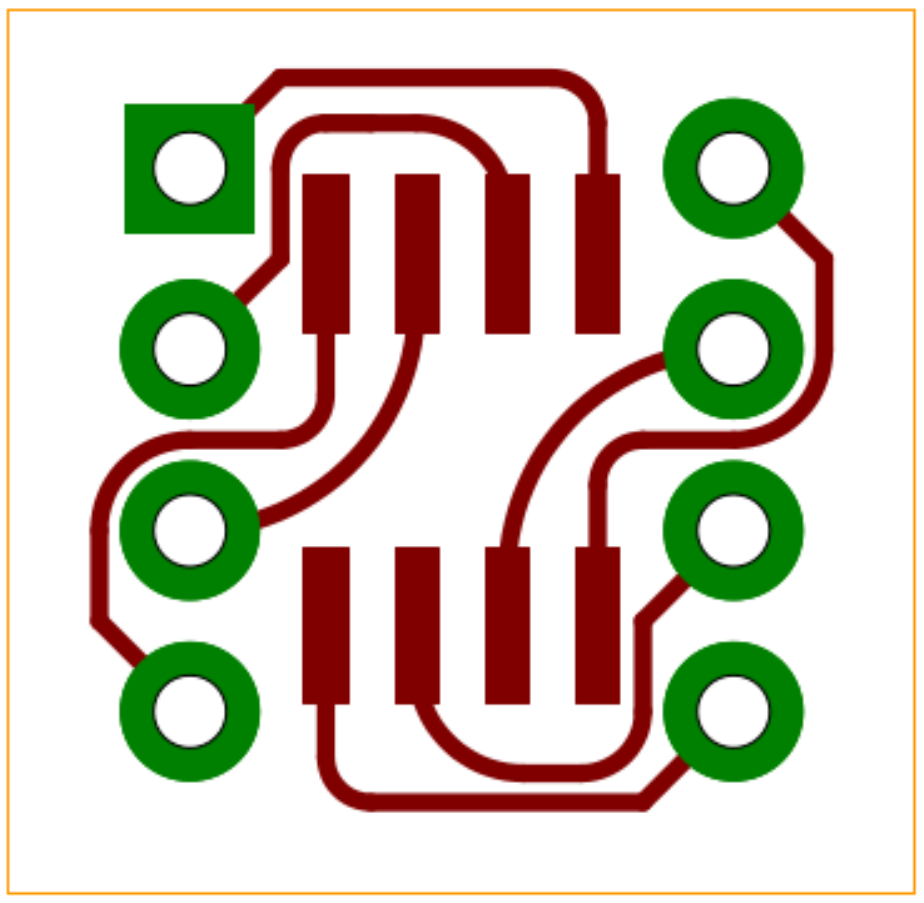}
\caption{Schematic representation of the SOIC-to-DIP adapter, shown in Fig.~\ref{fig:soic-to-dip} and Fig.~\ref{fig:4898}, where the operational amplifier is soldered on.}
\label{fig:StoD}
\end{figure}
\begin{figure}[h]
\includegraphics[scale=0.33]{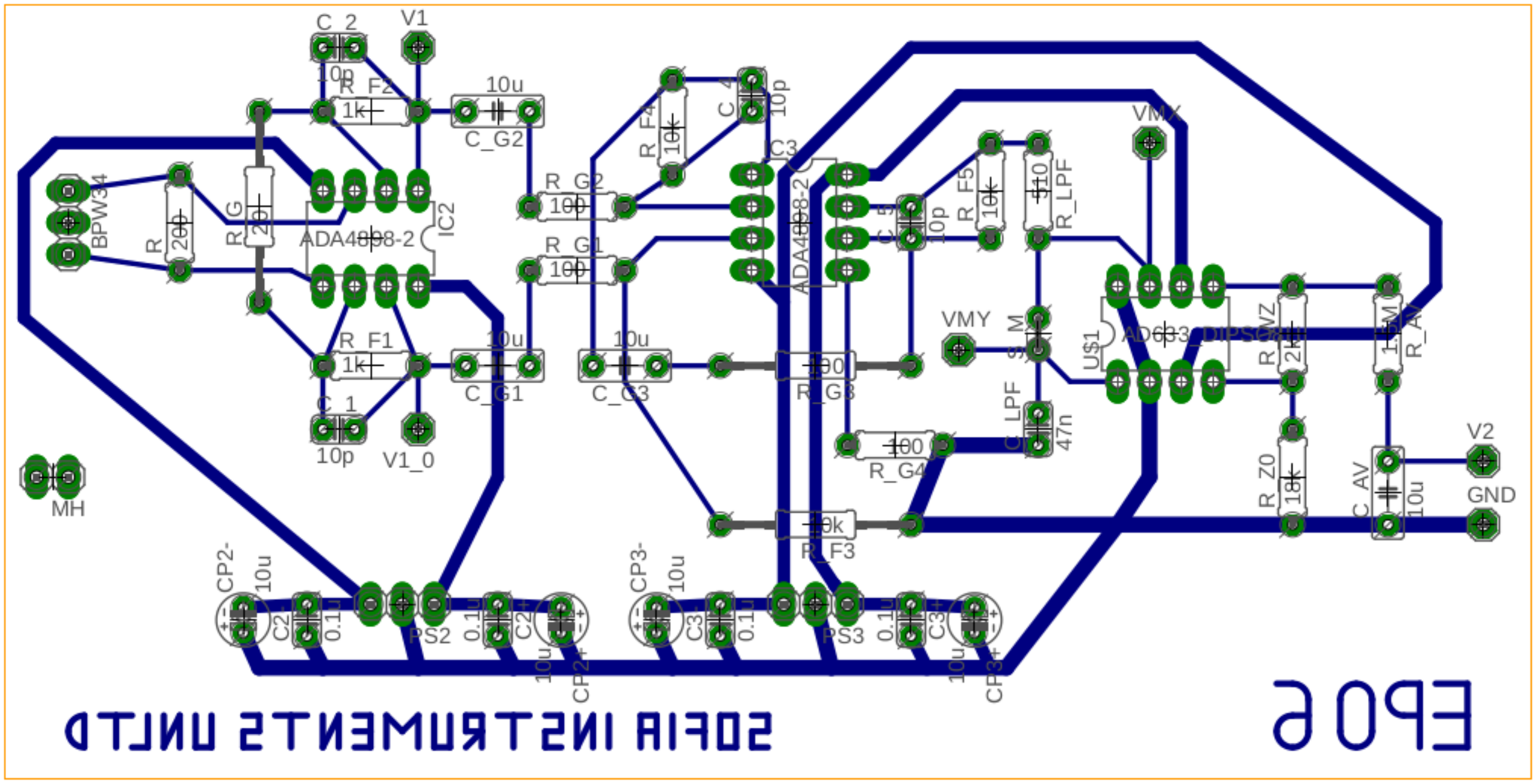}
\caption{Schematic representation of the PCB board of the experimental set-up, shown in Fig.~\ref{fig:photo_set-up}.}
\label{fig:board}
\end{figure}

\end{document}